\documentclass[twocolumn,
showpacs, preprintnumbers, amsmath, amssymb, nodata]{revtex4}
\usepackage{graphicx}
\usepackage{epsfig}
\def\Ref#1{(\ref{#1})}
\newcommand{\be}{\begin{equation}}
\newcommand{\ee}{\end{equation}}
\newcommand{\bn}{\begin{eqnarray}}
\newcommand{\en}{\end{eqnarray}}
\usepackage[english]{babel}

\begin{document}
\draft

\title{Analysis of the Dynamics of Liquid Aluminium: Recurrent
Relation Approach}

\author{A. V. Mokshin$^{1,2}$, R. M. Yulmetyev$^{1,2}$, R. M. Khusnutdinoff$^{1,2}$ and P. H\"anggi$^3$}

\address{$^1$Department of
Physics, Kazan State University, Kazan, Kremlyovskaya 18, 420008,
Russia}

\address{$^2$Department of Physics, Kazan State Pedagogical
University, Kazan, Mezhlauk 1, 420021,  Russia}

\address{$^3$Department of Physics, University of Augsburg,
Augsburg, Universit\"atsstr. 1, D-86135  Germany}

\date{\today}

\begin{abstract}
By use of the  recurrent relation approach (RRA)  we study the
microscopic dynamics of liquid aluminium at $T=973$\/ K and
develop a theoretical model which satisfies all the corresponding
sum rules. The investigation covers the inelastic features as well
as the crossover of our theory into the hydrodynamical and the
free-particle regimes. A comparison between our theoretical
results with those following from a generalized hydrodynamical
approach is also presented. In addition to this we report the
results of our molecular dynamics simulations for liquid
aluminium, which are also discussed and compared to experimental
data. The received results reveal that (i) the microscopical
dynamics of density fluctuations is defined mainly by the first
four even frequency moments of the dynamic structure factor, and
(ii) the inherent relation of the high-frequency collective
excitations observed in experimental spectra of dynamic structure
factor $S(k,\omega)$ with the two-, three- and four-particle
correlations.
\end{abstract}



\maketitle

\section{\label{intr} Introduction}

The study of dynamical processes related to atomic motions in
liquid systems present an attractive research topic, both of
theoretical and experimental origin. This being so, the extent of
corresponding experimental knowledge about microscopic processes
in liquids has  continuously increased and revealed new features
for various classes of liquids, which in turn call for an
appropriate theoretical explanation \cite{Tullio_RMP}. As an
intriguing example  we  mention the well-defined oscillatory modes
observed in liquid metals occurring outside the hydrodynamic
region and clearly detectable as corresponding high-frequency
peaks in the experimentally relevant observable -- the dynamic
structure factor $S(k,\omega)$
\cite{Boon,Barrat,March,Hansen,Balucani_Zoppi}. Thus, spectra of
the dynamic structure factor can display, even on microscopical
spatial scales, the triple-peak form known in the hydrodynamics as
the `Brillouin triplet'.

At present various theories do exist (see, for instance,
references
\cite{McGreevy,Bruin,Gotze,Schepper,Tullio1,Bafile,Bove,Hosokawa}),
which suggest an explanation of the `Brillouin triplet' evolving
at the finite values of wave-number $k$. Their description may be
found in excellent reviews \cite{Tullio_RMP,Yoshida,Copley}, and
we can assess that a great number of these theories are based on
the generalization of equations of linearized hydrodynamics.
However, it is necessary to point out that hydrodynamical
equations are a physically correct description for a large
wavelengths and time scales, where slow processes are strongly
pronounced and consequently provide the significant contribution.
Therefore, any wave-number and especially \emph{time}
(\emph{frequency}) extension of various thermodynamic and
transport quantities or the inclusion of the additional
non-hydrodynamical contributions should be invoked with some care.
A need to do so is particularly necessary for the description of
fast processes, which are a major component of microscopic
dynamics of liquids and in fact  may contribute appreciably to the
short-time behavior of corresponding relaxation functions (and/or
into high-frequency scenario of their power spectra). The
existence of finite frequency moments  of $S(k,\omega)$ can serve
as a peculiar criterion to test and estimate the validity of a
corresponding  theoretical extension \cite{Lee1}.

In the present work we study the microscopic dynamics of liquid
aluminium on the basis of the theory developed within the
framework of the so-called \emph{recurrent relation approach}
(RRA) \cite{Lee1}. According to the RRA, the time evolution of a
dynamical variable depends only on basis vectors, which span the
underlying Hilbert space of the studied dynamical variable and the
dimensionality of the embedding space $\mathcal{S}$ \cite{Lee2}.
Relaxation functions and terms defining the  interactions appear
here as projections of dynamical variables on the corresponding
orthogonal basis vectors. If the dynamical variable is a density
fluctuation then the RRA allows one to obtain the corresponding
relaxation function in terms of frequency moments of the dynamic
structure factor.

As the specific physical system under consideration we choose the
dynamics of liquid aluminium. This choice is related, foremost, to
the availability of high precision data  for the Inelastic X-ray
Scattering (IXS)  for  liquid aluminium near its melting point
\cite{Tullio_Al}. These data are \emph{generally} consistent with
the results of molecular dynamics simulations with various models
for the inter-particle potentials \cite{Gonz,Gonz2,Ebs}.
Nevertheless, it is necessary to note that the full agreement
between experimental and simulation methods in characterization of
dynamical features of liquid aluminium does not achieved (see
p.$916$ of Ref. \cite{Tullio_RMP}).

The organization of the paper is as follows: In the next Section
we present the theoretical fundamentals; in particular we review
within our context the basic notions of the RRA
\cite{Lee1,Lee2,Lee3} as these apply to the density fluctuation
case considered in the given work. In section \ref{reg} the study
of various dynamical regimes and their features for the case of
liquid aluminium is presented. We also report in this section
details and results of our molecular dynamics simulations, which
are compared with IXS-data. A resume with some concluding remarks
is given in section \ref{remar}.

\section{\label{sec:level1} Theoretical framework}

Let us consider a system of $N$ identical classical particles of
 mass $m$ evolving in the volume $V$. We  take the density fluctuations
\begin{equation}
A^{(0)}(\textbf{k})=\frac{1}{\sqrt{N}}\sum_{j=1}^{N} \textrm{e}^{i
\textbf{k} \cdot \textbf{r}_j} \label{W0}
\end{equation}
as an `initial' dynamical variable; $\textbf{k}$ is the
wave-vector. The time evolution of $A^{(0)}(\textbf{k})$ is
defined by the Heisenberg equation
\begin{equation}
\frac{dA^{(0)}(\textbf{k},t)}{dt}=i[\widehat{\mathcal{H}},A^{(0)}(\textbf{k},t)]
=i\hat{\mathcal{L}}A^{(0)}(\textbf{k},t), \label{eq_m}
\end{equation}
where $\hat{\mathcal{L}}$ is the Liouville operator,
which is Hermitian, $\widehat{\mathcal{H}}$ is the Hamiltonian of
the system and $[\widehat{\mathcal{H}},\ldots]$ is the Poisson
bracket. According to the RRA scheme \cite{Lee1,Lee2} the quantity
$A^{(0)}(\textbf{k},t)$ can be considered as a vector in an
\textit{a priori} chosen  $d$-dimensional space $\mathcal{S}$,
i.e.
\begin{eqnarray}
A^{(0)}(k,t)=\sum_{\nu=0}^{d-1}a_{\nu}^{(0)}(k,t)f_{\nu}(k),\ \ \
k=|\textbf{k}|\textrm{ is fixed}. \label{RRA1}
\end{eqnarray}
 Here the `basis
vectors' $f_0(k)$, $f_1(k)$, ..., $f_{d-1}(k)$ spanning
$\mathcal{S}$ are orthogonal, i.e.,
\begin{eqnarray}
(f_{\nu}(k),f_{\mu}(k))=(f_{\nu}(k),f_{\nu}(k)) \delta_{\nu\mu},
\end{eqnarray}
and are interrelated by the following recurrent relation (RR-I):
\begin{eqnarray}
f_{\nu+1}(k)&=&i\hat{\mathcal{L}}f_{\nu}(k)
+\Delta_{\nu}(k)f_{\nu-1}(k), \ \ \
 \nu\geq0,\nonumber\\
\Delta_{\nu}(k)&=&\frac{(f_{\nu}(k),f_{\nu}(k))}{(f_{\nu-1}(k),f_{\nu-1}(k))},\nonumber\\
\ \ f_{-1}(k)&=&0,\ \ \ \ \Delta_{0}(k)\equiv1. \label{RR1}
\end{eqnarray}
Here $a_{\nu}^{(0)}(k,t)$ is the time dependent projection of
$A^{(0)}(k,t)$ on the $\nu$-th basis vector $f_{\nu}(k)$; the
brackets $(...,...)$ denotes the  scalar product of Kubo in the
embedding space $\mathcal{S}$ \cite{Lee1}. Then, the relaxation
function between two variables $X$ and $Y$ is defined by
\begin{equation}
(X,Y)= \frac{1}{\beta} \int_0^{\beta} \langle \exp(\lambda
\widehat{\mathcal{H}})Y^\dag \exp(-\lambda \widehat{\mathcal{H}})X
\rangle d\lambda,
\end{equation}
where $X,~Y \subset \mathcal{S}$, $\beta=(k_B T)^{-1}$, $k_B$ and
$T$ are the Boltzmann constant and temperature, respectively
\cite{Kubo}. The angular brackets denote the average over the
canonical ensemble with temperature $T=(k_B \beta)^{-1}$. Note
that in the classical case ($\beta \to 0$, $\hbar \to 0$), the
relaxation function proportional to the usual correlation function
\cite{Lee1}
\begin{equation}
(X,Y) \equiv \langle X Y^* \rangle. \label{crossov}
\end{equation}

On the basis of the set of vectors $\{f_{n}(k)\}$ one can
construct a set of dynamical variables $\{\textbf{A}^{(n)}(k)\}$
which we need for the description of the evolution of the system,
obeying
\begin{eqnarray}
 A^{(n)}(k,t=0)=f_{n}(k), \ \ \ n=0,\ 1,\ ...,\ d-1.
\label{dyn_var}
\end{eqnarray}
The corresponding time evolution by analogy with equation
\Ref{RRA1} can be defined as
\begin{eqnarray}
A^{(n)}(k,t)=\sum_{\nu=n}^{d-1}a_{\nu}^{(n)}(k,t)f_{\nu}(k).
\label{RRAG}
\end{eqnarray}
Note that the set of functions $a_{n}^{(n)}(k,t)$ (i.e. for
$n=\nu$) in equations \Ref{RRA1} and \Ref{RRAG} are defined by
projecting the dynamical variable $A^{(n)}(k,t)$ onto the
corresponding basis $f_n(k)$, i.e.
\begin{eqnarray}
a_{n}^{(n)}(k,t)&=&\frac{(A^{(n)}(k,t),f_{n}(k))}{(f_{n}(k),f_{n}(k))}\nonumber\\
&=&
\frac{(A^{(n)}(k,t),A^{(n)}(k,0))}{(A^{(n)}(k,0),A^{(n)}(k,0))},
\label{rel_fun}
\end{eqnarray}
where the last equality is obtained by taking into account
equation \Ref{dyn_var}. So, one can easily see that these
functions are \emph{normalized}, i.e.
\begin{eqnarray}
a_{n}^{(n)}(k,t=0)=1. \label{fir}
\end{eqnarray}
Here and afterwards  we simplify notation whenever the quantity
involves the same upper and lower index by setting:
$a^{(n)}(k,t)=a_{n}^{(n)}(k,t)$.

The functions $a_{\nu}^{(n)}(k,t)$ for $\nu>n$, being defined by
the projection of the vector $A^{(n)}(k,t)$ on the basis
$f_{\nu}(k)$ with $n\neq \nu$, are related to  interactions
between $A^{(n)}(k)$ and $A^{(\nu)}(k)$. Further, the set
$a_{\nu}^{(n)}(k,t)$ obeys the following properties:
\begin{eqnarray}
\textrm{if \ \ \ }\nu>n \textrm{\ \ \ then \ \ \ } & &
a_{\nu}^{(n)}(k,t=0)=0, \label{oth} \\
\textrm{if \ \ \ }\nu<n \textrm{\ \ \ then \ \ \ } &
&a_{\nu}^{(n)}(k,t)\equiv 0. \nonumber
\end{eqnarray}

It is useful to note that the functions $a_{\nu}^{(n)}(k,t)$ are
interrelated by the second recurrent relation (RR-II) \cite{Lee1}, reading:
\begin{eqnarray}
\label{RR2} \Delta_{\nu+1}(k) a_{\nu+1}^{(n)}(k,t)
 = -\frac{da_{\nu}^{(n)}(k,t)}{dt}
+a_{\nu-1}^{(n)}(k,t),
\\
n = 0,\ 1,\ 2,\ \ldots \ d-1, \nonumber\\
\nu = n,\ n+1,\ \ldots \ d-1, \nonumber\\
\nu \geq n. \nonumber
\end{eqnarray}
Thus, the set of recurrent relations \Ref{RRA1} and \Ref{RR2}
yield the time evolution of the dynamical variable $A^{(n)}(k)$,
which occurs in the space spanned by the orthogonal basis vectors
$f_n(k)$, $f_{n+1}(k)$, ... .

Similar to the well-known  Zwanzig-Mori formalism, these recurrent relations obey
the  equation \cite{Lee3}:
\begin{eqnarray}
\frac{d}{dt}A^{(n)}(k,t)&=&A^{(n+1)}(k,t)- \Delta_{n+1}(k)\\
& &\times \int_0^t a^{(n+1)}(k,t')A^{(n)}(k,t-t')dt', \nonumber
\end{eqnarray}
which constitutes  the exact reformulation of equation \Ref{eq_m}.
The last equation can be rewritten in terms of the Laplace
transform $\widetilde{f}(z)=\int_0^{\infty}\textrm{e}^{-zt}f(t)dt$
of the functions $a^{(n)}(t)$ as:
\begin{eqnarray}
\widetilde{a}^{(n)}(k,z)=[z+\Delta_{n+1}(k)\widetilde{a}^{(n+1)}(k,z)]^{-1}.
\label{step}
\end{eqnarray}
Then, the Laplace transform of relaxation function of the density
fluctuation $\widetilde{a}^{(0)}(k,z)$ can be represented in the
form of a continued fraction, i.e., \cite{Mori,Scof,Scof_Tem}
\begin{equation}
\widetilde{a}^{(0)}(k,z)= \frac{1}{\displaystyle
z+\frac{\Delta_{1}(k)}{\displaystyle
z+\frac{\Delta_{2}(k)}{\displaystyle
z+\frac{\Delta_{3}(k)}{z+\ddots}}}}. \label{fraction}
\end{equation}

Thus, the time behavior of density relaxation function
$a^{(0)}(k,t)$ is fully defined by the  dimensionality $d$ of the
space $\mathcal{S}$ and by the corresponding set of parameters
$\Delta_{n+1}(k)$ ($n=0$, $1$, $\ldots$). To determine
$A^{(0)}(k,t)$ [and $a^{(0)}(k,t)$] it is necessary to know the
whole set of the parameters $\Delta_{n+1}(k)$.

The parameters $\Delta_{n+1}(k)$ can be expressed in terms of
\emph{normalized} frequency moments of the dynamical structure
factor $S(k,\omega)$ \cite{Balucani_Zoppi,l1}, i.e.,
\begin{eqnarray}
\omega^{(p)}(k)&=&\left.
(-i)^p\frac{d^{p}a^{(0)}(k,t)}{dt^p}\right |_{t=0} \nonumber\\&=&
\frac{\int_{-\infty}^{\infty}\omega^{p}
S(k,\omega)d\omega}{\int_{-\infty}^{\infty}S(k,\omega)d\omega},
\label{fr_m}
\end{eqnarray}
which is related, in turn, with $a^{(0)}(k,t)$ through the Fourier
transform as
\begin{eqnarray}
S(k,\omega)=\frac{S(k)}{2\pi}\int_{-\infty}^{\infty}\textrm{e}^{i\omega
t} a^{(0)}(k,t) dt.
\end{eqnarray}
Because $S(k,\omega)$ is even function of $\omega$ for a classical
system, all odd frequency moments are equal to zero. Moreover,
from equations \Ref{RR2} and \Ref{fr_m} and taking into account
equations \Ref{fir} and \Ref{oth} one finds the relations
\begin{eqnarray}
 \label{param}
\omega^{(2)}(k)&=&\Delta_1(k),\\  
\omega^{(4)}(k)&=&\Delta_1^2(k)+\Delta_1(k) \Delta_2(k),\nonumber\\
\omega^{(6)}(k)&=&\Delta_1(k)[\Delta_1(k)+\Delta_2(k)]^2
+\Delta_1(k)\Delta_2(k)
\Delta_3(k),\nonumber\\
\omega^{(8)}(k)&=&\Delta_1(k) \{[\Delta_1(k)+\Delta_2(k)]^3+
2\Delta_2(k)\Delta_3(k)\nonumber\\& &\ \ \ \times
[\Delta_1(k)+\Delta_2(k)]+
\Delta_2(k)\Delta_3^2(k)\}\nonumber\\
& &+\Delta_1(k)\Delta_2(k)\Delta_3(k)\Delta_4(k),\nonumber\\
\omega^{(10)}(k)&=&\Delta_1(k) \{\Delta_1(k)[\Delta_1(k) +\Delta_2(k)]^3\nonumber\\
& &+\Delta_1(k)\Delta_2(k)
[\Delta_1(k)+\Delta_2(k)]\nonumber\\
& &\ \ \ \times[\Delta_1(k)+\Delta_2(k)+\Delta_3(k)]\nonumber\\
& &+ \Delta_2(k)\Delta_3(k)
[\Delta_1(k)+\Delta_2(k)+\Delta_3(k)]\nonumber\\
& &\ \ \ \times[\Delta_1(k)+\Delta_2(k)
+\Delta_3(k)+\Delta_4(k)] \nonumber\\
&
&+\Delta_2^2(k)[\Delta_1(k)+\Delta_2(k)+\Delta_3(k)]^2\nonumber\\
& &+ \Delta_1(k)\Delta_2(k)\Delta_3(k)[\Delta_1(k)+\Delta_2(k)]\nonumber\\
&
&+\Delta_2(k)\Delta_3(k)\Delta_4(k)[\Delta_1(k)+\Delta_2(k)\nonumber\\
& &\ \ \ \ \ \ \ \ \ \ \ \ \ \ \ \ \ \
 +\Delta_3(k)+\Delta_4(k)+\Delta_5(k)]\}, \nonumber\\
  & & \ldots\ .  \nonumber
\end{eqnarray}

\noindent On the other hand, in the classical case, the frequency
parameters $\Delta_{n}(k)$ can be obtained in accordance with the
help of equations \Ref{RR1} and \Ref{crossov} to read: 
\label{coef}
\begin{eqnarray}
 \label{fr_c}
\Delta_{1}(k)&=& \frac{\langle|f_{1}|^2 \rangle}{\langle|f_{0}|^2
\rangle} = \frac{k_BT}{m}\frac{k^2}{S(k)}, \\
\langle|f_{1}|^2 \rangle &=& \frac{k_BT}{m} k^2,~ \langle|f_{0}|^2
\rangle = S(k), \nonumber
\end{eqnarray}

By analogy with equation \Ref{fr_c} it follows that
\begin{eqnarray}
\Delta_{2}(k)&=&  \frac{k_BT}{m}k^2  \left (
3-\frac{1}{S(k)} \right ) \label{sec_c} \\
&+& \frac{\rho}{m} \int \nabla_l^2 u(r)[1-
\exp(i\textbf{k}\cdot\textbf{r})]g(r) d^3\textbf{r}, \nonumber
\end{eqnarray}
\begin{eqnarray}
\label{f_c} \Delta_{3}(k)&=& \frac{1}{\Delta_{2}(k)} \left \{ 15
\left ( \frac{k_BT}{m}k^2 \right )^2 +
\mathcal{F}(k)  \right \} \nonumber\\
&-&\frac{1}{\Delta_{2}(k)}\left [\Delta_{1}(k)+\Delta_{2}(k)
\right ]^2.
\end{eqnarray}
Here, $\rho$ denotes the number density, $S(k)$ is the static
structure factor, $g(r)$ is the pair distribution function, $u(r)$
is the inter-particle potential and the suffix $l$ denotes the
component parallel to $\textbf{k}$, whereas the term
$\mathcal{F}(k)$ denotes the combination of integral expressions
containing the inter-particle potential with two- and
three-particle distribution functions. In the general case, the
parameters $\Delta_{n}(k)$ at large $n$th order also contain the
distribution functions of $n$, $(n-1)$, ... and $n=2$.
 As a result, we can see that if the studied system is
characterized by strongly pronounced potential interactions then
the problem of finding  the time (frequency) dependence of the
dynamical variables $A^{(n)}(k)$ and/or the functions
$a^{(n)}(k,t)$ [$\widetilde{a}^{(n)}(k,z)$, see equations
\Ref{step} and \Ref{fraction}] is reduced to the  problem of
truncating the chain of coupled $n$-th particle distribution
functions \cite{Bogoliubov}.

\section{Dynamical regimes and their features} \label{reg}
\subsection{Short wavelength dynamics}
Let us consider the spatio-temporal regime of a mono-atomic
liquid, for which one can neglect the interaction between
particles. Obviously, this case corresponds to the short-time
dynamics which is restricted to  length scales smaller than the
mean free path (i.e., the regime of high $k$-values). Then, the
terms characterizing the strength of interaction between particles
are negligible, and the static structure factor $S(k) \to 1$. Upon
observing that the second (integral) term in equation \Ref{sec_c}
and $\mathcal{F}(k)$ in equation \Ref{f_c} are negligible one
finds
\begin{eqnarray}
 \Delta_1(k)&=&\frac{k_BT}{m}k^2,\nonumber\\
\Delta_2(k)&=&2\Delta_1(k),\ \ \Delta_3(k)=3\Delta_1(k).
\label{gauss_rec}
\end{eqnarray}
From the first equality of equation \Ref{fr_m} and equation
\Ref{param} one derives that equations \Ref{gauss_rec} represent
the short-time behavior of $a^{(0)}(k,t)$ described by the
Gaussian function,
\begin{eqnarray}
a^{(0)}(k,t)=\textrm{e}^{-\Delta_1(k)t^2/2}. \label{gauss}
\end{eqnarray}
Then, from equation \Ref{RR2} one finds that other functions
$a_{n}^{(0)}(k,t)$ defining interactions of density fluctuations
with other dynamical variables assume the following form
\begin{eqnarray}
a_{n}^{(0)}(k,t)=\frac{t^n}{n!}\textrm{e}^{-\Delta_1(k)t^2/2}, \ \
n=0,\ 1,\ \ldots \ .
\end{eqnarray}

Note in this case that the values of time scales determined by
$\sim 1/\Delta_{n+1}(k)$ with the increase of $n$ become smaller,
and the ratio between neighboring terms converges from $2$ towards $1$ upon increasing
 $n$; explicitly we have
 \begin{eqnarray}
\frac{\Delta_{n+2}(k)}{\Delta_{n+1}(k)} =\frac{n+2}{n+1}.
\label{rel}
\end{eqnarray}
As a result, for this regime of large $k$-values (opposite to the
hydrodynamic limit) one can conclude that although the time scales
of relaxation functions upon increasing of $n$ also increase, they
eventually become equal, i.e $\lim_{n \to \infty}
\Delta_{n+2}(k)/\Delta_{n+1}(k)=1$.

\subsection{The fluid dynamics at intermediate wavelengths}

Obviously, the simplifications used in the previous case are
incorrect for microscopic spatial ranges on the scale of order of
some inter-particle distances, where interaction between particles
is considerable. This implies a more complex numerical
(theoretical) evaluation of the parameters $\Delta_{n+1}(k)$,
starting with the second parameter, i.e at $n=1$ and higher.

On the other hand, the parameters $\Delta_{n}(k)$ can be found
from molecular dynamics simulations. This is very convenient and
useful tool, but it requires $\emph{a priori}$ the detailed
information about the interaction of particles in the study
system. According to this method, one can find the 'initial'
dynamical variable $A^{(0)}(k)$ from simulation data [see equation
\Ref{W0}], and then receive numerically the structural parameters
$\Delta_{n}(k)$ from RR-I [see equation \Ref{RR1}]. Obviously, for
higher precision the carried out results should be averaged over
time iterations. \noindent
\begin{table*}
\caption{Values of the static structure factor $S(k)$, the first
unnormalized frequency moment $\omega_{un}^{(2)}(k)$ and the
scattering intensity at the zeroth frequency $I(k,\omega=0)$ for
liquid aluminium. IXS denotes results of reference
\cite{Tullio_Al} obtained from Inelastic X-ray Scattering data,
whereas MD indicates outcomes of our molecular dynamics
simulations. \label{md_exp}}
\begin{ruledtabular}
\begin{tabular}{c|ccccccc} &\multicolumn{2}{c}{$S(k)$} &
\multicolumn{3}{c}{$\omega_{un}^{(2)}(k)$} &
\multicolumn{2}{c}{$I(k,\omega=0)$~($10^{-3}$ ps)} \\  \cline{2-3}
\cline{4-6} \cline{7-8}  $k$ (nm$^{-1}$) & IXS & MD & $k_BTk^2/m$
& IXS & MD & IXS & MD
\\ \hline
4.2  & 0.010 & 0.029 & 5.288  & 5.294  &  5.443  & 0.250 & 0.805 \\
5.4  & 0.013 & 0.032 & 8.741  & 8.732  &  11.010 & 0.320 & 0.826 \\
7.8  & 0.016 & 0.034 & 18.239 & 18.241 &  23.170 & 0.350 & 0.837 \\
9.0  & 0.020 & 0.037 & 24.283 & 24.192 &  32.758 & 0.390 & 0.844 \\
10.2  & 0.025 & 0.037 & 31.190 & 30.925 &  40.351 & 0.425 & 0.896 \\
11.4  & 0.027 & 0.045 & 38.961 & 39.021 &  47.598 & 0.434 & 0.910 \\
12.6  & 0.030 & 0.050 & 47.595 & 47.631 &  59.238 & 0.520 & 1.174 \\
13.8  & 0.036 & 0.059 & 57.092 & 57.137 &  72.048 & 0.601 & 1.253 \\
\end{tabular}
\end{ruledtabular}
\end{table*}

\begin{figure}
\hspace{2cm} {\epsfig{figure=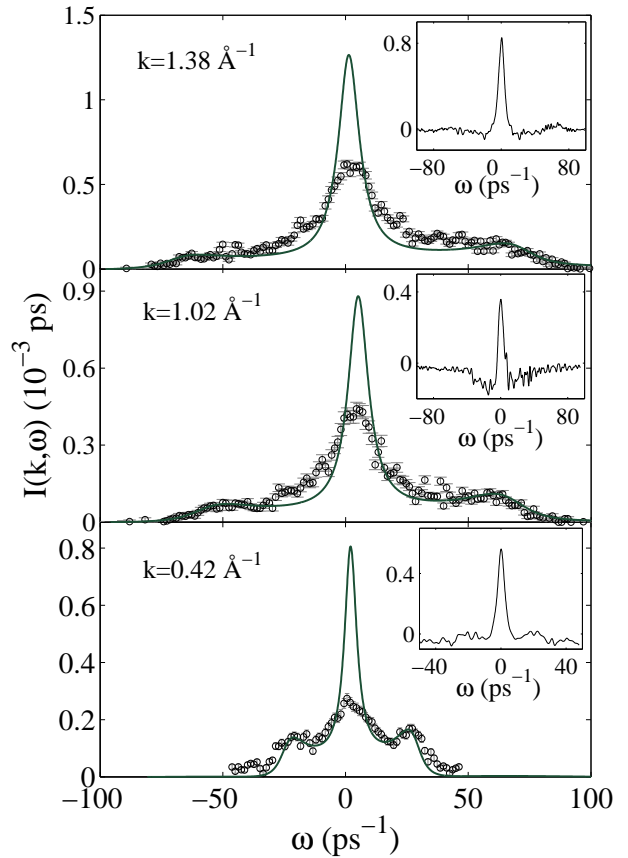,height=12cm,angle=0} }
\caption{Main plots: the scattering intensity $I(k,\omega)$ of
liquid aluminium at temperature $T=973$~K for different
wave-numbers. The solid lines present the molecular dynamics
results convoluted with the experimental resolution and involving
the detailed balance condition, open circles with error bars are
the IXS-data of reference \cite{Tullio_Al}. Insets: the difference
between molecular dynamics results and interpolated experimental
data at the fixed $k$, i.e. $I_{MD}(k,\omega)-I_{IXS}(k,\omega)$,
in the units $10^{-3}$ps.
 \label{Ikw_md}}
\end{figure}
\noindent We have performed such numerical analysis based on our
molecular dynamics simulations for liquid aluminium with the
particle density $n=0.0528$ $\textrm{\AA}^{-3}$ at the temperature
$T=1000$ K. The interaction of $N=4000$ particles imbedded in the
cubic cell ($L=42.32$ $\textrm{\AA}$) with the periodic boundary
conditions was realized using the so-called `glue' potential
\cite{Erc}. The time step $\Delta t$ applied in the integration of
the equation of motion was $10^{-14}$ s. After the bringing of the
system to equilibrium state, $100000$ time steps were made
\cite{Dib,l2}.

Results for the static structure factor $S(k) = \langle | f_0(k)
|\rangle$ and the first \emph{unnormalized} frequency moment
$\omega_{un}^{(1)}(k) = \langle | f_1(k) |\rangle $ deduced from
molecular dynamics simulations are presented in table
\ref{md_exp}, where these quantities are compared with the
corresponding values $S(k) = \int S(k,\omega) d\omega$ and
$\omega_{un}^{(1)}(k) = \int \omega^2 S(k,\omega) d\omega$
obtained on the basis of experimental IXS-data for the dynamical
structure factor $S(k,\omega)$ of liquid aluminium
\cite{Tullio_Al}. In this table we also give the exact theoretical
prescription for the first unnormalized frequency moment
$\omega_{un}^{(2)}(k)=k_B T/m$.

As can be seen, the theoretical values of $\omega_{un}^{(2)}(k)$
are in agreement with the IXS-data of reference \cite{Tullio_Al}
that is a quite expected, because the normalization of
experimental data is performed to the first sum rules (for
details, see also reference \cite{Tullio_RMP}), whereas the values
of $\omega_{un}^{(2)}(k)$ obtained from simulations are
overestimated in comparison to IXS-data for all wave-numbers. The
values of the static structure factor $S(k)$ from molecular
dynamics are also higher of IXS-data ($S^{MD}(k)/S^{IXS}(k) \sim
1.5 - 2.9$). As a result, the ratio between the values of the
first structural parameter $\Delta_{1}(k)=\langle | f_1(k)
|\rangle / S(k)$ from IXS-data \cite{l3} and the corresponding
values from molecular dynamics simulations is $\sim 1.2-2.8$. The
analysis performed by us reveals that the difference of high-order
structural parameters obtained from IXS-data and from molecular
dynamics simulations is also observed, namely,
$\Delta_2^{MD}(k)/\Delta_2^{IXS}(k) \sim 2.2 - 6.7$ and
$\Delta_3^{MD}(k)/\Delta_3^{IXS}(k) \sim 1.8 - 6.1$. Figure
\ref{Ikw_md} compares the experimental scattering intensity
$I(k,\omega)$ of liquid aluminium near melting temperature
\cite{Tullio_Al} and the results of molecular dynamics
simulations. Although results of simulations are in a good
agreement with the IXS-data for the high-frequency region,
simulation data give higher values of central peak in comparison
to the experimental ones. The exact values of $I(k,\omega=0)$ from
numerical simulations and from IXS are presented in table
\ref{md_exp}. On this basis, one can conclude that although the
`glue' potential of reference \cite{Erc} describes correctly some
equilibrium characteristics of liquid aluminium, it is unsuitable
to reproduce the microscopical dynamics features of liquid
aluminium, and it is unfit for finding of the structural
parameters $\Delta_n(k)$.

There exists also another scheme which is based on a
phenomenological estimation of the structural parameters
$\Delta_n(k)$. The recent experimental IXS-data of the intensity
$I(k,\omega)$ for liquid aluminium at $T=973$~K \cite{Tullio_Al}
allows one to estimate the experimental frequency moments and, as
a result, to identify $\Delta_{n+1}(k)$. For this procedure we
employed as a fitting function the model for the classical dynamic
structure factor $S(k,\omega)$ as detailed in \cite{Tankeshwar},
i.e.,
\begin{eqnarray}
 S(k,\omega)&=&S(k)\frac{a \tau_1}{2}
\textrm{sech}\left ( \frac{\pi \omega \tau_1}{2} \right
)\nonumber\\ &+& S(k)\frac{(1-a)\tau_2}{4}\left [
\textrm{sech}\left (\frac{\pi (\omega+\omega_0) \tau_2}{2} \right
) \right. \nonumber\\ & &\ \ \ \ \ + \left. \textrm{sech}\left
(\frac{\pi (\omega-\omega_0) \tau_2}{2} \right ) \right ]\;,
\end{eqnarray}
with the model parameters $a$, $\tau_1$, $\tau_2$ and $\omega_0$.
It is worthwhile to note that a similar approach has been used
also in \cite{Larsson}, where the experimental $S(k,\omega)$ of
liquid metals was fitted by a sum of three Gaussian functions in
order to derive the first and second order memory functions
numerically. However, as it was established during the fitting
procedure for liquid aluminium, the model function of
\cite{Tankeshwar} allows one to realize a better adjustment to the
experimental data (see figure 1) than the combination of Gaussian
functions. As a result, after taking into account the detailed
balance condition \cite{Tullio_RMP,Lee2}, i.e.,
\begin{equation}
S_{q}(k,\omega) = \frac{\hbar \omega/k_B T}{1-\textrm{e}^{-\hbar
\omega/k_B T}} S(k, \omega), \label{dbcond.}
\end{equation}
the convolution of $S_{q}(k,\omega)$ with the resolution function
$R(k,\omega)$ was deduced to read:
\begin{equation}
I(k,\omega)=E(k)\int R(k,\omega-\omega')S_q(k,\omega') d\omega,
\label{resol}
\end{equation}
wherein $E(k)$ denotes the form factor. The resulting function
$I_{f}(k,\omega)$ has been adjusted to the experimental scattering
function so that the  the first two sum rules are obeyed
identically [for $S(k)$ and $\Delta_1(k)$]. The subscript $f$ used
here for the intensity $I(k,\omega)$ denotes the fact that it is
the result of a fitting procedure. The comparison of the
theoretical normalized second frequency parameter from equation
\Ref{fr_c} with the obtained one from integration of the fitting
dynamic structure factor $S(k,\omega)$ according to equation
\Ref{fr_m} is depicted in the main part of figure \ref{om_Al}.
Finally, the remaining parameters $\Delta_{n+1}(k)$ ($n=1$, $2$,
$\ldots$) were deduced from $S(k,\omega)$, which yields the best
fitting of $I_{f}(k,\omega)$ to the experimental data. The found
values of the first five parameters \emph{vs.} $k$ are given in
the inset of figure \ref{om_Al}.

\begin{figure*}
\centerline {\epsfig{figure=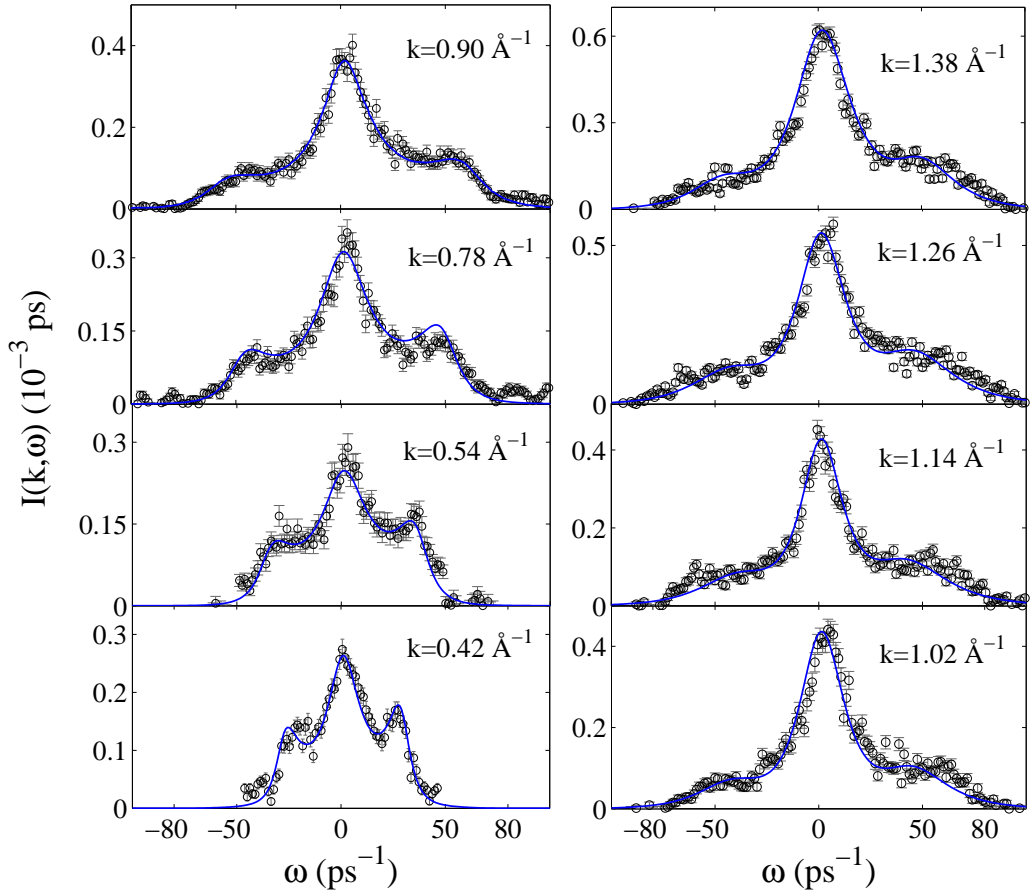,height=22cm,angle=-90}}
\caption{\label{Skw_Al}  The scattering intensity $I(k,\omega)$ of
liquid aluminium at temperature $T=973$ K. The solid lines depict
the results of the theoretical model \Ref{Basic}, whereas the open
circles present the IXS-data of reference \cite{Tullio_Al}. The
theoretical line-shapes have been modified to account for the
quantum mechanical detailed balance condition and have  been
broadened for the finite experimental resolution effects as
described in the text.}
\end{figure*}

\begin{figure*}
\hspace{1cm} {\epsfig{figure=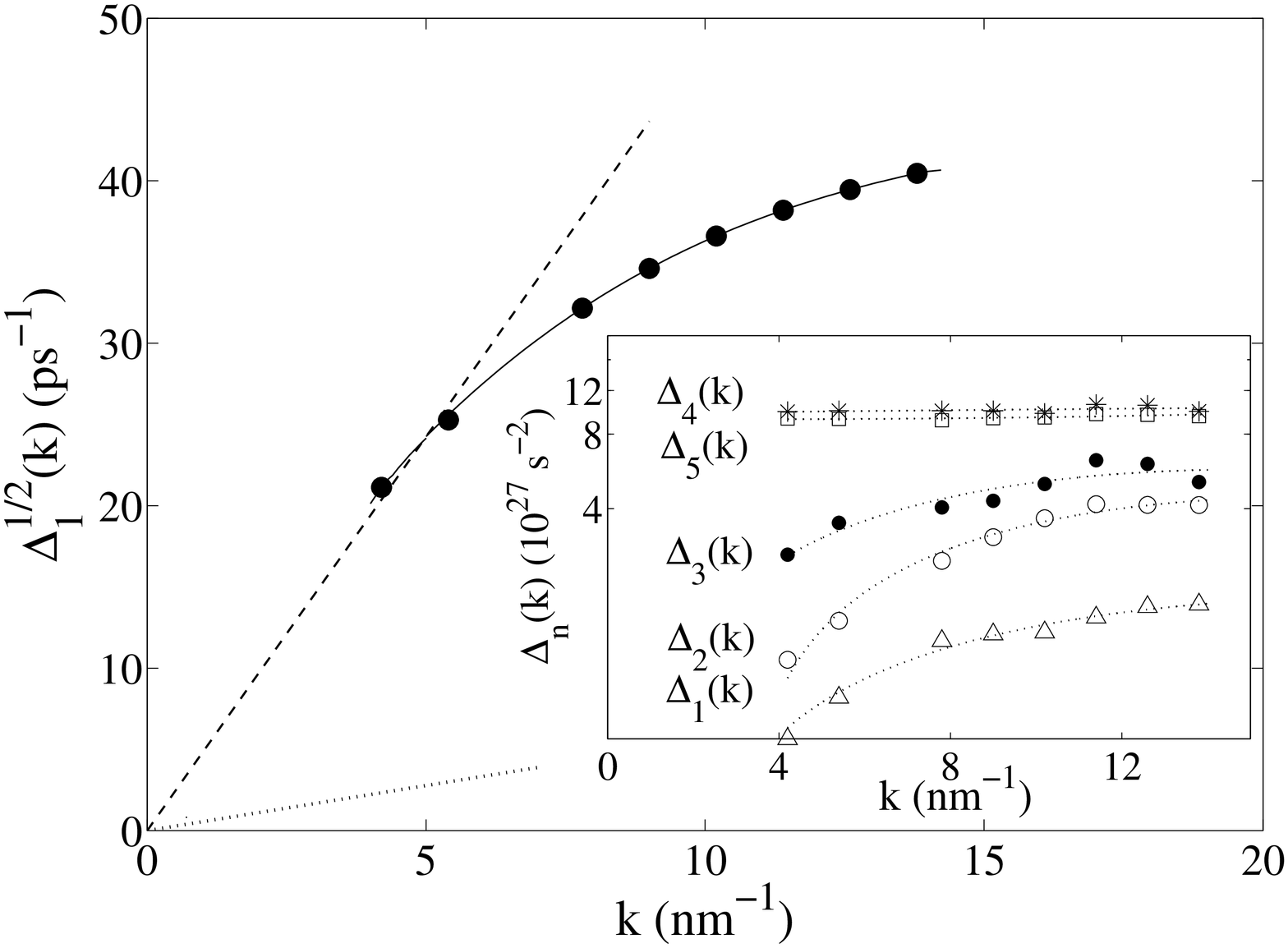,height=14cm,angle=-90}}
\caption{\label{om_Al} Main: the solid line corresponds to the
theoretical results of the second frequency moment from equation
\Ref{fr_c}, whereas the full circles present the values extracted
by means of an integration of the resolution de-convoluted,
classical $S(k,\omega)$ used to reproduce the experimental
IXS-data (see in the text). The dashed and the dotted lines
present the hydrodynamical limit (see equation \Ref{hydr_om1} with
the isothermal sound velocity $c_0(k)=4700$~m/s, from reference
\cite{Tullio_Al}) and the free-particle situations
($\Delta_1^{1/2}(k)=1/\sqrt{\beta m}\; k$), respectively. Inset:
the values of the first five parameters obtained from the
frequency moments of the dynamic structure factor $S(k,\omega)$
corresponding to the best fit of $I_f(k,\omega)$ to the
experimental data, as detailed in the text.}
\end{figure*}

During this procedure the following features for the studied
wave-number region were identified:

(i) although the frequency moments and parameters
$\Delta_{n+1}(k)$ are sensitive to the form of $S(k,\omega)$, the
ratio of the neighboring parameters,
$\delta_{n}(k)=\Delta_{n+1}(k)/\Delta_{n}(k)$ ($n=1$, $2$,
$\ldots$), is invariable for the various fitted functions, which
reproduce the experimental data within the acceptable bounds;

(ii) with increasing $n$ for the studied wave-number region the
following feature of these parameters is observed:
$\Delta_{n+1}(k) > \Delta_{n}(k)$ ($\delta_n(k)>1$). This feature
is not valid for  $n=4$;

(iii) $\Delta_4(k)$ is \emph{slightly larger} than $\Delta_5(k)$,
$0.915 \leq \delta_4(k) \leq 0.964$ (see table \ref{tab1}).

\begin{table}
\caption{\label{tab1} Ratio of neighboring parameters
$\delta_n(k)=\Delta_{n+1}(k)/\Delta_{n}(k)$ ($n=1$, $2$, $3$ and
$4$) deduced according to equation \Ref{param} on the basis of
experimental frequency moments for liquid aluminium at $T=973$~K.}
\begin{ruledtabular}
\begin{tabular}{ccccc}
 $k$ $(\textrm{nm}^{-1})$ & $\delta_1(k)$ &
 $\delta_2(k)$ &
 $\delta_3(k)$ &
 $\delta_4(k)$ \\
\hline
4.2  & 2.008 & 1.357 & 3.787 & 0.935  \\
5.4  & 2.043 & 1.484 & 2.839 & 0.926  \\
7.8  & 2.105 & 1.344 & 2.455 & 0.915  \\
9.0  & 2.466 & 1.301 & 2.310 & 0.934  \\
10.2 & 2.890 & 1.376 & 1.926 & 0.964  \\
11.4 & 2.866 & 1.504 & 1.680 & 0.920  \\
12.6 & 2.581 & 1.465 & 1.723 & 0.920  \\
13.8 & 2.519 & 1.239 & 1.929 & 0.957  \\
\end{tabular}
\end{ruledtabular}
\end{table}
\noindent On this basis, we further suppose that all  ratios of
higher order beginning with $n=4$ are approximately equal to $1$,
i.e. $\delta_{n}(k)\simeq 1$ for $n \geq 4$. This suffices  to
obtain the exact equation for dynamic structure factor. From the
physical point of view, taking into account that the frequency
parameters can be considered as characteristics of time scales for
corresponding quantities \cite{Lee2,AVM2}, such an approximation
means that the average time scales of the relaxation processes
related to the dynamical variables $A^{(n)}$ ($n\geq4$) are
approximately equal, i.e.
$\Delta_n^{-1/2}(k)=\Delta_{n+1}^{-1/2}(k)$ for $n\geq4$.
Therefore, the time scales of `neighbor' relaxation processes in
this spatial range become equal at low levels and become strongly
pronounced in comparison with the short wavelength dynamics. As a
result, in the framework of the RRA one can exactly obtain the
relaxation function $a_{3}^{(3)}(k,t)$ in the following form:
\begin{eqnarray}
a^{(3)}(k,t)=\frac{1}{\sqrt{\Delta_4(k)}t}J_1(2\sqrt{\Delta_4(k)}t),
\label{bessel}
\end{eqnarray}
where $J_1$ is the Bessel function of the first order. Further,
relaxation functions of other dynamical variables and functions
characterizing their interactions can be expressed by means of
integral equations. However, the term of the greatest interest
\textit{a priori} is $a^{(0)}(k,t)$, the Fourier transform of
which, $S(k, \omega)$, presents a experimentally measurable
quantity. Therefore, from equation \Ref{bessel} we obtain
\begin{eqnarray}
\widetilde{a}^{(3)}(k,z)=\frac{-z+[z^2+4\Delta_4(k)]^{1/2}}{2\Delta_4(k)}
\;.
\end{eqnarray}
Then, from equations \Ref{fraction} and \Ref{RR2} we can deduce
the dynamic structure factor to read:
\begin{widetext}
\bn \label{Basic} 2 \pi \frac{S(k, \omega)}{S(k)}&=&
\frac{\Delta_{1}(k) \Delta_{2}(k)
\Delta_{3}(k)}{\Delta_4(k)-\Delta_3(k)} \frac{[4 \Delta_{4}(k)-
\omega^{2}]^{1/2}}{\omega^6 +\mathcal{A}_1(k) \omega^4
+\mathcal{A}_2(k) \omega^2 +\mathcal{A}_3(k)},\\
\mathcal{A}_1(k)&=&
\frac{\Delta_3^2(k)-\Delta_2(k)[2\Delta_4(k)-\Delta_3(k)]}{\Delta_4(k)-\Delta_3(k)}
-2\Delta_1(k),\nonumber\\
\mathcal{A}_2(k)&=&\frac{\Delta_2^2(k)\Delta_4(k)-2\Delta_1(k)\Delta_3^2(k)+\Delta_1(k)\Delta_2(k)
[2\Delta_4(k)-\Delta_3(k)]}{\Delta_4(k)-\Delta_3(k)}+\Delta_1^2(k),\nonumber\\
\mathcal{A}_3(k)&=&\frac{\Delta_1^2(k)\Delta_3^2(k)}{\Delta_4(k)-\Delta_3(k)}.
\nonumber \en
\end{widetext}

\noindent Note that the denominator of the well known hydrodynamic
dynamic structure factor $S(k,\omega)$ is also represented as the
bicubic polynomial (in the variable $\omega$) of equation
\Ref{Basic} (see reference \cite{link1}), whereas the numerator of
the hydrodynamic model contains the biquadratic polynomial. The
point is that at the crossover into the hydrodynamical region (as
well as into the region of free-particle motion) the equation for
the dynamic structure factor \Ref{Basic} becomes modified. This is
so because the rules of ratio for the frequency parameters are
also changed. As will be presented in the next section, the so
developed theory yields the correct hydrodynamical asymptotic
behavior.

The comparison of the theoretical results of the dynamic structure
factor $S(k,\omega)$ of liquid aluminium at $T=973$ K according to
equation \Ref{Basic} with data of IXS are presented with figure
\ref{Skw_Al}. As can be deduced from this figure, the theoretical
results reproduce well the experimental data.

\subsection{High-frequency modes}

From the analysis of the experiments it is known that inelastic
features of scattering spectra are pronounced in the normalized
longitudinal current relaxation function
\begin{eqnarray}
\label{CCF} G_J(k,t) =
\frac{(J^L(k,0),J^L(k,t))}{(J^L(k,0),J^L(k,0))},
\end{eqnarray}
which is related to the dynamic structure factor by
\begin{eqnarray}
S(k) \Delta_1(k) \widetilde{G}_J(k,\omega)= \omega^2 S(k,\omega).
\end{eqnarray}
The last equation can be readily obtained from
\begin{eqnarray}
\Delta_1(k) G_J(k,t) = -\frac{\partial^2 a^{(0)}(k,t)}{\partial
t^2}. \label{cur_rel}
\end{eqnarray}
The quantity $\widetilde{G}_J(k,\omega)$ possesses one minimum at
$\omega=0$ and two high-frequency maxima, being  contrary to the
dynamic structure factor. Moreover, the side peaks in
$\widetilde{G}_J(k,\omega)$ correspond to those of $S(k,\omega)$.

The inelastic features of the spectrum $\widetilde{G}_J(k,\omega)$
are defined by the solution in regard to $z=z(k)$ of the following
equation (see equation (2.54) in Ref. \cite{Yoshida}):
\begin{eqnarray}
z+\frac{\Delta_1(k)}{z}+\Delta_2(k)\widetilde{a}^{(2)}(k,z)=0,
\label{cond}
\end{eqnarray}
where $\widetilde{a}^{(2)}(k,z)$, according to the above obtained
results, has the form
\begin{eqnarray}
\widetilde{a}^{(2)}(k,z)=\frac{2\Delta_4(k)}{z(2\Delta_4(k)-
\Delta_3(k)) +\Delta_3(k)\sqrt{z^2+ 4\Delta_4(k)}}. \nonumber \\
\label{a_2}
\end{eqnarray}
Generally, equation \Ref{cond} possesses complex solutions
$z=\textrm{Re}[z(k)]+i\textrm{Im}[z(k)]$, where
$\textrm{Im}[z(k)]$ defines the positions of the inelastic peaks
in $\widetilde{G}_J(k,\omega)$, whereas $\textrm{Re}[z(k)]$
characterizes the widths of these peaks.

Introducing for convenience the  notation $\mathcal{Q}(k)$ for the
ratio between the parameters $\Delta_4(k)$ and $\Delta_3(k)$, and
the dimensionless quantity $\xi(k)$, with the latter
characterizing the frequency region:
\begin{subequations}
\bn
\mathcal{Q}(k)=2\frac{\Delta_4(k)}{\Delta_3(k)} - 1, \label{Q}
\en
\bn
\xi(k)=\frac{z^2}{\Delta_4(k)}, \label{small_p}
\en
\end{subequations}
the condition for the existence of side peaks in Eq. \Ref{cond},
can be rewritten in the following form (see Ref.
\cite{link2,Zwanzig,RUBIN}): \bn z^2+z\Delta_2(k)\frac{1
+\mathcal{Q}(k)}{z\mathcal{Q}(k) +\sqrt{z^2+ 4\Delta_4(k)}}
+\Delta_1(k)=0. \label{cond_fin} \en

To investigate this  equation further it is convenient to consider the following
limiting situations. First, the region of crossover into the
hydrodynamical limit can be characterized by $|\xi(k)| \ll 1$.
This condition allows one to span the region of small frequencies
(i.e. the large time scales). Then, the dispersion equation takes the
following form:
\begin{eqnarray}
\label{cub} z^3+ \frac{2\Delta_4^{1/2}(k)}{\mathcal{Q}(k)}z^2
&+&\left [\Delta_1(k)+
\frac{\Delta_2(k)(1+\mathcal{Q}(k))}{\mathcal{Q}(k)}\right
]z\nonumber\\ &+&
\frac{2\Delta_4^{1/2}(k)\Delta_1(k)}{\mathcal{Q}(k)}=0.
\end{eqnarray}
Although the exact algebraic solution of a cubic equation is
feasible, it is not of much practical use here  because its
inherent algebraic complexity. Let us note that equation \Ref{cub}
is similar to the dispersion equation obtained by Mountain (see
equation (15) in reference \cite{Mountain}). Therefore, following
the convergent scheme for approximating solutions \cite{Mountain}
one finds
\begin{eqnarray}
\label{disper} z_{1,2}(k)&=&\pm ic_sk
-\Gamma k^2,\\
z_{3}(k)&=&-2\frac{\Delta_4^{1/2}(k)}{\gamma
\mathcal{Q}(k)},\nonumber
\end{eqnarray}
where the adiabatic sound velocity $c_s$, the sound damping
parameter $\Gamma$, and the ratio of the specific heat at constant
pressure to the specific heat at constant volume $\gamma=c_p/c_v$
read for $k \to 0$: \begin{subequations}
\begin{eqnarray}
c_s&=&\sqrt{\gamma} c_{0}, \ \ \ \  \lim_{k \to 0}\Delta_1(k) = c_{0}^2  k^2, \label{hydr_om1}\\
\Gamma&=&\frac{\gamma-1}{\gamma}
\frac{\Delta_4^{1/2}(k)}{\mathcal{Q}(k)},\\
\gamma&=&1+\frac{\Delta_2(k)[1+\mathcal{Q}(k)]}
{\Delta_1(k)\mathcal{Q}(k)},
\end{eqnarray}
\end{subequations}
respectively, and $c_0(k)$ is the isothermal sound velocity. The
equations \Ref{disper} are the results of the hydrodynamical
Landau-Placzek theory \cite{Landau}, and the real and imaginary
parts of the first two solutions define the positions and widths
of the Mandelshtam-Brillouin doublet. The approximate solutions
\Ref{disper} are valid, when the difference between $\Delta_4(k)$
and $\Delta_3(k)$ is sufficiently large in comparison with
differences between $\Delta_3(k)$, $\Delta_2(k)$ and $\Delta_1(k)$
\cite{note1,sod}. This condition corresponds to the requirement of
the Mountain approximation procedure (p.$208$, \cite{Mountain})
and is related to the divergence of the frequency parameters in
the hydrodynamical limit.

If one considers the large-frequency region (regardless of $k$),
i.e. the opposite extreme with $z^2/\Delta_4(k) \gg 1$, then here
equation \Ref{cond_fin} yields
\begin{eqnarray}
z_{1,2}(k) \approx \pm i \sqrt{\Delta_1(k)+ \Delta_2(k)} \equiv
\pm i \omega_L(k), \label{solid}
\end{eqnarray}
which corresponds to a typical `instantaneous' solid-like response
\cite{Balucani_Zoppi,l4}.

\begin{figure}
\centerline{\epsfig{figure=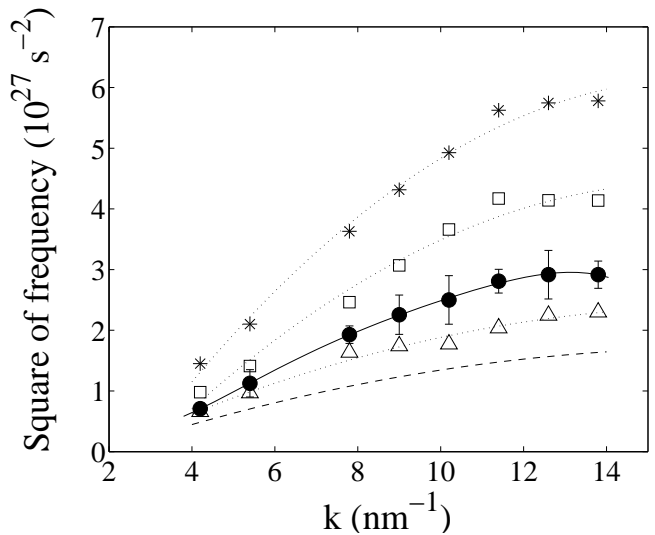,height=7cm,angle=0}}
\caption{\label{disp} The short dashed line corresponds to
$\Delta_1(k)$ from equation \Ref{fr_c}; the open triangles present
the values of $\gamma \Delta_1(k)$, which possesses the limiting
behavior $\lim_{k \to 0} \gamma \Delta_1(k)= (c_s k)^2$ and thus
corresponds to the squared frequency of the Brillouin peak from
hydrodynamical prescription ($\gamma=c_p/c_v=1.4$ from references
\cite{Inui,Tullio_RMP}); the filled circles denote the squared
frequency of the side peak $\omega_c^2(k)$ in the experimental
IXS-spectra \cite{Tullio_Al}; the solid line presents the
theoretical predictions; the open squares are the values of
$\Delta_2(k)$, which are obtained through the second and fourth
normalized frequency moments of the resolution de-convoluted,
classical dynamic structure function $S(k,\omega)$ extracted from
the  IXS-spectra [see equation \Ref{param}]. The squared frequency
of the `solid-like' excitations
$\Delta_1(k)+\Delta_2(k)\equiv\omega_L^2(k)$ is presented by the
stars.}
\end{figure}

As can be seen from the dispersion equation \Ref{cond_fin} both,
the widths and positions of the side peaks depend on the first
four parameters $\Delta_n(k)$ ($n=1$, $2$, $3$ and $4$). In figure
\ref{disp} the square of various frequencies are compared to the
square of the frequency of collective excitations,
$\omega_c^2(k)$, extracted from the experimental data
\cite{Tullio_Al}. The results of figure \ref{disp} justify that
the values of the squared frequencies of the side peaks are
arranged into the intermediate region between `hydrodynamical
prescriptions' with $\gamma \Delta_1(k)$ and values
$\omega_L^2(k)$ corresponding to the `solid-like' behavior
[equation \Ref{solid}], while the results of our model yield the
true squared frequencies of side peak, $\omega_c^2(k)$. The fact,
that inelastic peaks are fully defined by the first four frequency
parameters, indicates directly that the existence of side peak
features in the studied region depend on both,  two- and
three-particle as well as four-particle interactions, which are
taken into account by these parameters. Although the
high-frequency  motions can be considered as a remnant of solid
dynamics, the large values of lifetimes of the underlying
excitations (in comparison with the solid-state ones) are
distinctive for the dynamics of a liquid, where two-, three- and
four-particle correlations are strongly pronounced at the length
scales of the order of some Angstroms.

\section{Resume and concluding remarks} \label{remar}

As mentioned in the Introduction, the treatment of the
experimental data of Inelastic Neutron Scattering (INS) and IXS
experiments can be conveniently performed in the framework of
generalized hydrodynamics methods. Recall that for the
hydrodynamical regime the following equation holds \cite{Copley}
\begin{eqnarray}
\label{hydr_m2} \Delta_2(k)
a^{(2)}(k,t)&=&\frac{2\eta_{L}k^{2}}{\rho m} \delta(t)+(\gamma
-1)\Delta_{1}(k)\textrm{e}^{-D_{T}k^{2}t}, \nonumber\\
\end{eqnarray}
where $\eta_{L}$ is the longitudinal viscosity and $D_{T}$ is the
thermal diffusion. The last equation is obtained from a reasoning
that is related to the hydrodynamical approach and is correct only
for the  dynamical description for large temporal and spatial
scales, i.e $t \to \infty$ and $k \to 0$. However, this form of
relaxation function breaks down when addressing the  short-time
features. From the point of view of the RRA the relaxation
function corresponding to the Hermitian system can not
\textit{exactly} assume the time dependence of the form
\Ref{hydr_m2}, because the normalized condition \Ref{fir} is
violated. One of the simplest generalization of the hydrodynamical
approach consists in the assumption about the extended (not
instantaneous) character of the decaying viscosity term involving
various elementary contributions. So, in references
\cite{Tullio_RMP,Tullio_Al} it was shown that for the correct
reproduction of the experimental $S(k,\omega)$ (including the case
of fluid aluminium \cite{Tullio_Al}) the \emph{three exponential
decay terms}  must be included at least in $a^{(2)}(k,t)$,
\begin{eqnarray}
a^{(2)}(k,t)=\sum_{j=\alpha,\mu,th} \mathcal{G}_{j}(k)
\textrm{e}^{-t/\tau_{j}(k)}, \label{visc}
\end{eqnarray}
where $\tau_{\alpha}(k)$ and $\tau_{\mu}(k)$ are the relaxation
times associated with viscosity processes, whereas
$\tau_{th}(k)=1/D_Tk^2$ is the relaxation time scale of thermal
contribution [the second term in equation \Ref{hydr_m2}],
$\mathcal{G}_{\alpha,\mu,th}$ are the weights of the corresponding
contributions. Although this \textit{ansatz} also has the
violation of the normalized condition \Ref{fir}, it can be used as
a sufficiently good approximation to study $THz$ frequency ranges
(see in reference \cite{AVM1}).

Indeed, as can be seen from figure \ref{om_Al}, the frequency
parameter $\Delta_4(k)$ assumes the most significant values (in
comparison to others) and thereby it defines the shortest time
scales $2\pi/\sqrt{\Delta_4(k)} \sim  10^{-14}$s. Then the
condition $|\xi(k)|\ll 1$ (see equation \ref{small_p}) allows one
to span the frequency (time) range $\omega < 10^{14}$s$^{-1}$
($t>10^{-14}$s), which corresponds to the microscopic dynamics and
is mainly available to IXS and INS. As a result, we find from
equation \Ref{a_2} that \cite{l5}
\begin{eqnarray}
\widetilde{a}^{(2)}(k,z)&=& \sum_j
\frac{\mathcal{G}_j(k)}{z+\tau_j^{-1}(k)}, \ \ \sum_j
\mathcal{G}_j(k) =1, \nonumber\\
& &\ \ \ \ \ j=1,\; 2,\; 3,\; 5,\; \ldots\; ,
\end{eqnarray}
where $\mathcal{G}_j(k)$ and $\tau_j(k)$ are expressed in terms of
$\Delta_3(k)$ and $\Delta_4(k)$ (see for details reference
\cite{AVM1}). The given equation at $j=3$  yields the same result
as equation \Ref{visc}, whereas the coarse approximation at $j=1$
corresponds to the  \emph{viscoelastic model}. Thus, we thereby
justify that the expansion of the function $a^{(2)}(k,t)$ into
elementary exponential contributions can be used as a good
approximation in the description of the  dynamics (but only) at
\emph{long} and \emph{intermediate} time scales \cite{Gotze}
bounded by the picosecond regime.

With this present study we have explored the  dynamics of liquid
aluminium at $T=973$ K on the basis of the RRA, when the
relaxation functions of the density fluctuations can be exactly
extracted from the known relations between the frequency
parameters $\delta_n(k)=\Delta_{n+1}(k)/\Delta_n(k)$, ($n=1$, $2$,
$\ldots$). In the formulation of the RRA the  parameters
$\Delta_n(k)$'s constitute the structural characteristics of the
embedding Hilbert space $\mathcal{S}$ of the investigated
dynamical variables. As a result, the peculiarities of the
dynamics depend directly on the dimension $d$ of the space
$\mathcal{S}$ and the related parameters, $\Delta_n(k)$'s. If, in
particular, $\delta_n(k) \to 0$, then the realized embedding space
$\mathcal{S}$ can be considered as the ($n+1$)dimensional one, and
the exact solution for the relaxation function can be found
\cite{Lee1,Lee2,Lee3}. In  the opposite situation $\delta_n(k) \to
\infty$, which  can be observed in the hydrodynamical limit, the
corresponding \emph{exact} solution causes problems due to the
divergence of the frequency parameters $\Delta_n(k)$ and
$\Delta_{n+1}(k)$.

In this work we extracted phenomenologically these frequency
parameters on the basis of the first five frequency moments of the
experimental dynamic structure factor. The values of these five
parameters increase with growing index $n$. A similar behavior
occurs for large values of $k$ (the free-particle case) where
these parameters approach each other at large $n$. In the studied
wave-number region this approximate identity is observed at $n=4$
already. The assumption that this holds as well at higher order
$n$ allows one to develop a theoretical model, which turns out to
be in quantitative agreement with the experimentally observed
IXS-data for liquid aluminium and which satisfies all the
corresponding sum rules. We found that the first four even
frequency moments appeared in the continued fraction of RRA (see
equation \Ref{fraction}) as well as in the Zwanzig-Mori's
formalism are necessary to restore the genuine microscopic
dynamics of liquid aluminium. Moreover, our analysis reveals that
the high-frequency features of microscopic collective dynamics in
liquid aluminium depend on two-, three- and four-particle
correlations. The given finding can be useful in detailed
investigations of equilibrium features of fluid aluminium, at the
estimation of the many-particle distribution functions, and in the
studying of cluster phenomena of liquid aluminium \cite{Noya} near
the melting.

\section{Acknowledgments} We thank M. H. Lee for helpful discussions and are
grateful to T. Scopigno for providing IXS-data of liquid
aluminium. This work is supported by Grant of RFBR No.
05-02-16639a, RNP No. 2.1.1.741 and by the German Research
Foundation, SFB-486, Project A10 (P.H.).


\begin{thebibliography}{10}

\bibitem{Tullio_RMP} T. Scopigno,
G. Ruocco and F. Sette, Rev. Mod. Phys  \textbf{77}, 881 (2005).

\bibitem{Boon}  J. P. Boon, and S. Yip, \textit{Molecular
Hydrodynamics} (McGraw-Hill, New York, 1980).

\bibitem{Barrat} J.-L. Barrat and J.-P. Hansen, \textit{Basic concepts for simple and complex liquids}
(University Press, Cambridge,  2003).

\bibitem{March} N. March, and M. Tosi, \textit{Atomic Dynamics in
Liquids} (Dover, New York, 1991).

\bibitem{Hansen} J.-P. Hansen, and I. McDonald, \textit{Theory of Simple
Liquids} (Academic, New York, 1986).

\bibitem{Balucani_Zoppi} U. Balucani and M. Zoppi, {\it Dynamics of the Liquid State}
(Clarendon Press, Oxford, 1994).

\bibitem{McGreevy} R. L. McGreevy and E. W. J. Mitchell, Phys. Rev. Lett.
\textbf{55}, 398 (1985).

\bibitem{Bruin} C. Bruin, J. C. van Rijs, L. A. de Graaf, and I. M. de Schepper,
Phys. Lett. A \textbf{110}, 40 (1988); I. M. de Schepper, E. G. D.
Cohen, C. Bruin, J. C. van Rijs, W. Montfrooij, L. A. de Graaf,
Phys. Rev. A \textbf{38}, 271 (1988).

\bibitem{Gotze} W. G\"otze, J. Phys. C \textbf{11}, A1 (1999); H.Z.
Cummins, ibid. \textbf{11}, A99 (1999); L. Sjogren, Phys. Rev. A
\textbf{22}, 2866 (1980); \textbf{22}, 2883 (1980); W. G\"otze and
M.R. Mayr, Phys. Rev. E \textbf{61}, 587 (2000).

\bibitem{Schepper} I. M. de Schepper, E. G. D. Cohen, C. Bruin,
J. C. van Rijs, W. Montfrooij, and L. A. de Graaf, Phys. Rev. A
\textbf{38}, 271 (1988); I. M. Mryglod, I. P. Omelyan, and M. V.
Tokarchuk, Mol. Phys. \textbf{84}, 235 (1995); T. Bryk and I.
Mryglod, Phys. Rev. E \textbf{63}, 051202 (2001).

\bibitem{Tullio1} T. Scopigno, U. Balucani, G. Ruocco, and F.
Sette, Phys. Rev. E \textbf{65}, 031205 (2002).

\bibitem{Bafile} U. Bafile, E. Guarini and F. Barocchi, Phys.
Rev. E \textbf{73}, 061203 (2006).

\bibitem{Bove} L. E. Bove, F. Sacchetti, C. Petrillo, and B.
Dorner, Phys. Rev. Lett. \textbf{85}, 5352 (2000); L. E. Bove, F.
Sacchetti, C. Petrillo, B. Dorner, F. Formisano, and F. Barocchi,
Phys. Rev. Lett. \textbf{87}, 215504 (2001); L.E. Bove, B. Dorner,
C. Petrillo, F. Sacchetti and J.-B. Suck, Phys. Rev. B
\textbf{68}, 024208 (2003); L.E. Bove, F. Formisano, F. Sacchetti,
C. Petrillo, A. Ivanov, B. Dorner and F. Barocchi, Phys. Rev. B
\textbf{71}, 014207 (2005).

\bibitem{Hosokawa} S. Hosokawa, Y. Kawakita, W.-C. Pilgrim, and H.
Sinn, Phys. Rev. B \textbf{63}, 134205 (2001).

\bibitem{Yoshida} F. Yoshida, S. Takeno, Phys. Rep. $\bf{173}$, 301 (1989).

\bibitem{Copley} J. R. D. Copley, S. W. Lovesey, Rep. Prog. Phys.
$\bf{38}$, 461 (1975).

\bibitem{Lee1} M. H. Lee, Phys. Rev. Lett. \textbf{51}, 1227
(1983); Phys. Rev. E \textbf{62}, 1769 (2000); Phys. Rev. E
\textbf{61}, 3571 (2000).

\bibitem{Lee2} U. Balucani, M. H. Lee, and V. Tognetti, Phys. Rep. \textbf{373},
409 (2003).

\bibitem{Tullio_Al} T. Scopigno, U. Balucani, G. Ruocco, and F.
Sette, Phys. Rev. E \textbf{63}, 011210 (2001).

\bibitem{Gonz} D. J. Gonz\'{a}lez, L. E. Gonz\'{a}lez, J. M.
L\'{o}pez, and M. J. Stott, Phys. Rev. B \textbf{65}, 184201
(2002); J. Chem. Phys. \textbf{115}, 2373 (2001).

\bibitem{Gonz2} D. J. Gonz\'{a}lez, L. E. Gonz\'{a}lez, and J. M. L\'{o}pez,
J. Phys. C \textbf{13}, 7801 (2001).

\bibitem{Ebs} I. Ebbsj\"{o}, T. Kinell, and I. Waller, J. Phys. C:
Solid St. Phys. \textbf{13}, 1865 (1980).

\bibitem{Lee3} M.H. Lee, Phys. Rev. E \textbf{62}, 1769 (2000),
M.H. Lee, Phys. Rev. Lett. \textbf{85}, 2422 (2000).

\bibitem{Mori} R. Zwanzig, Phys. Rev. \textbf{124}, 1338
(1961); H. Mori, Prog. Theor. Phys. \textbf{33}, 423 (1965); Prog.
Theor. Phys. \textbf{34}, 399 (1965).

\bibitem{Scof} P. Schofield, \textit{Inelastic Scattering of Neutrons in Solids
and Liquids} (IAEA, Vienna, 1961).

\bibitem{Scof_Tem}  P. Schofield, \textit{Physics of Simple Liquids}
ed. H.N.V. Temperley, J.S. Rowlinson and G.S. Rushbrooke
(Amsterdam, 1968).

\bibitem{l1} It is convenient sometimes to use the \emph{unnormalized}
frequency moments
$\omega^{(p)}_{un}(k)=\int_{-\infty}^{\infty}\omega^{p}
S(k,\omega)d\omega$. In this definition, the zeroth frequency
moment characterizes the static structure factor, i.e.
$\omega^{(0)}_{un}(k) \equiv S(k)$.

\bibitem{Bogoliubov} N. N. Bogoliubov, {\it Problems of Dynamic Theory in
Statistical Physics} (Gostekhizdat, Moscow-Leningrad, 1946) (in
Russian) [Reprinted in: Studies in statistical mechanics
\textbf{1} (J. de Boer and G. E. Uhlenbeck, eds., Amsterdam,
North-Holland, 1962)].

\bibitem{Erc} F. Ercolessi, M. Parrinello and E. Tosatti,
Philos. Mag. A \textbf{58}, 213 (1988); F. Ercolessi and J. B.
Adams, Europhys. Lett. \textbf{26}, 583 (1994).

\bibitem{Dib} J. J. Erpenberck and W. W. Wood, Phys. Rev.
A \textbf{26}, 1648 (1982).

\bibitem{l2} To estimate the frequency range allowable at the molecular
dynamics simulation for cell with this size \cite{Dib}, one can
find the inverse time required for a sound wave to cross the
entire periodic cell. At the length $L=42.32$ $\textrm{\AA}$ and
the sound velocity $c_s \simeq 4750$~m/s, it is possible to
consider the frequency range $\omega \geq c_s/L$, i.e $\omega \geq
1$~ps$^{-1}$.

\bibitem{l3} The procedure of the phenomenological estimation (from
IXS-data) of structural parameters will be discussed in detail
below.

\bibitem{Tankeshwar} S. Singh, and K. Tankeshwar, Phys. Rev. E
\textbf{67}, 012201 (2003).

\bibitem{Larsson} K.-E. Larsson, and W. Gudowski, Phys. Rev. A
\textbf{33}, 1968 (1986).

\bibitem{AVM2} A. V. Mokshin, R. M. Yulmetyev, and P. H\"anggi,
Phys. Rev. Lett. \textbf{95}, 200601 (2005).

\bibitem{link1} The comparison to result of hydrodynamical model
allows one to find the following correspondence:\\
$\mathcal{A}_1(k)=2[(\Gamma
k^2)^2-(c_0k)^2]+(D_Tk^2)^2, \\
\mathcal{A}_2(k)= [(\Gamma k^2)^2+(c_0k)^2]^2+ 2(D_Tk^2)^2[(\Gamma
k^2)^2-(c_0k)^2],\\ \mathcal{A}_3(k)= (D_Tk^2)^2 [(\Gamma k^2)^2
+(c_0k)^2]^2$.

\bibitem{Inui} M. Inui, S. Takeda, and T. Uechi, J. Phys. Soc. Jpn. \textbf{61}, 3203
(1992)

\bibitem{link2} It is interesting to note that the fraction in the second term of
Eq. \Ref{cond_fin} has the same form as the Laplace transform of
the velocity relaxation function of the  model by Rubin
\cite{Zwanzig,RUBIN} -- `impurity in the harmonic lattice', where
the ratio $\Delta_4(k)/\Delta_3(k)$ corresponds to the mass ratio
in the Rubin model.

\bibitem{Zwanzig} R. Zwanzig \textit{Nonequilibrium statistical
mechanics} (University Press, Oxford, 2001).

\bibitem{RUBIN}
R.~J.~Rubin, Phys. Rev. \textbf{131}, 964 (1963),  R.~J.~Rubin, J.
Amer.~Chem. Soc.  \textbf{90}, 3061 (1968); R.~J.~Rubin and
R.~Zwanzig, J.~Math.~Phys. \textbf{2}, 861 (1961).


\bibitem{Mountain} R.D. Mountain, Rev. Mod. Phys. \textbf{38}, 205
(1966).

\bibitem{l4} The same asympotic behavior is recovered from the viscoelastic
model (see for example reference \cite{Balucani_Zoppi}).

\bibitem{Landau} L. Landau and G. Placzek, Physik Z. Sowjetunion
\textbf{5}, 172 (1934).
\bibitem{note1} In reference \cite{sod} it was found that for the case of liquid
sodium the ratio $\Delta_4(k)/\Delta_3(k)$ assumes $100$ and even
higher at smaller, microscopic spatial scales.

\bibitem{sod} R. M. Yulmetyev, A. V. Mokshin, T. Scopigno, and P. H\"anggi,
J. Phys.: Condens. Matter \textbf{15}, 2235 (2003); R. Yulmetyev,
P. H\"anggi and F.~Gafarov, Phys. Rev. E \textbf{62}, 6178 (2000).

\bibitem{AVM1} A. V. Mokshin, R. M. Yulmetyev, and P. H\"anggi, J.
Chem. Phys. \textbf{121}, 7341 (2004).

\bibitem{l5} Note that expansion
(44) does not contradict to claims of RRA. It should be taken into
account, that this expansion is suitable for finite frequency
range only.

\bibitem{Noya}  E. G. Noya, J. P. K. Doye, and F. Calvo,
Phys. Rev. B \textbf{73}, 125407 (2006).

\end{thebibliography}
\end{document}